\documentclass[twocolumn,aps,amsmath,amssymb,prx,footinbib]{revtex4}
\usepackage{graphicx}
\usepackage{bm}         
\usepackage{color}         
\usepackage{amsmath}
\usepackage{units}
\usepackage{bbm}
\usepackage{hyperref}

\date{\today}
\begin{document}

\title{Conductance quantization in topological Josephson trijunctions}
\author{Julia~S.~Meyer and Manuel~Houzet}
\affiliation{Univ.~Grenoble Alpes, CEA, { Grenoble INP, IRIG, PHELIQS,} 38000 Grenoble, France}

\begin{abstract}
Majorana zero modes are attracting a lot of attention as a new kind of emergent quasiparticles in condensed matter physics. Here we propose a new and striking signature of the coupling between Majorana zero modes in a three-terminal Josephson junction made with topological superconductors. { Recently three and four-terminal junctions made either with conventional or with topological superconductors have been shown to exhibit Weyl crossings in their Andreev spectrum that lead to a non-zero quantized transconductance, like in the quantum Hall effect. In contrast with the other types of junctions, where the effect is limited to certain parameter regimes, we predict that, generically, {\em any} three-terminal Josephson junction made with topological superconductors is characterized by such a non-zero quantized transconductance.} The quantization value is $\pm2e^2/h$. Based on numerics, we check that this prediction is robust. 
\end{abstract}

\maketitle

\section{Introduction}

{
The Josephson current flowing in a junction between two {superconductors} is a striking manifestation of macroscopic quantum coherence, with applications in metrology and quantum information. This equilibrium current is related with the formation of Andreev {states localized in the junction,} whose energy depends periodically on the superconducting phase difference~\cite{Kulik1970,Beenakker1991,Pillet2010}. Topology emerged as a guide for {predicting exotic properties of Andreev states}. In particular, topological superconductors host Majorana modes at their ends~\cite{Kitaev2001,Lutchyn2010,Oreg2010,Mourik2012}. Then, in a junction with such leads, the hybridization of two Majorana modes results in an Andreev state with a period-doubling of its {energy-phase dependence}~\cite{Kitaev2001,4pi-e}. 
Furthermore, topologically protected crossings between Andreev states in junctions with more than two leads may be revealed through a quantized transconductance~\cite{Riwar2016,Yokoyama2015,Eriksson2017}. 
{The prediction motivated recent efforts to fabricate multi-terminal junctions~\cite{Cohen2018,Draelos2019,Pankratova2018,Graziano2019}}.
Here we combine {both topological} effects to predict a robust non-vanishing quantized  transconductance  in trijunctions with topological leads. Such devices are envisioned to reveal the anyonic nature of Majorana states through their braiding~\cite{Alicea2011,Aasen2016}. Our prediction {can be used to assess} that a given junction is indeed suitable to perform its braiding function. 
}

A series of recent works have explored the topological properties of Andreev states in multi-terminal Josephson junctions~\cite{Riwar2016,Yokoyama2015,Eriksson2017,Meyer2017,Levchenko2017,Deb2018,Houzet2019,Xie2019,Gavensky2019}. The findings exploit the analogy between the Andreev spectrum of an $N$-terminal junction as a function of the superconducting phase differences and the band structure of a material in $d=N\!-\!1$ dimensions, with the superconducting phase differences playing an analogous role as the quasimomenta in the Brillouin zone. 

A transition from a topologically trivial to a topologically non-trivial phase requires the closing of a gap. Therefore {the investigation of topological properties of multi-terminal Josephson junctions has mainly been focused on the identification} of topologically protected band crossings and, in particular, so-called Weyl points. Generically a crossing of two bands requires three control parameters. Thus, in a four-terminal junction, the three independent superconducting phase differences are sufficient to induce a topological transition~\cite{Riwar2016}, while in trijunctions an additional control parameter such as the flux through the junction area is required~\cite{Meyer2017,Levchenko2017}. In both cases, the appearance of Weyl points does not necessitate any fine tuning of the scattering properties of the junction. 

If we then consider the Andreev spectrum as a function of two phase differences, taking the third parameter as a tuning parameter, a Weyl point at zero energy signals a topological transition associated with a change of the Chern number {characterizing} the ground state of the effective two-dimensional system. As a consequence, upon fixing the tuning parameter, the effective two-dimensional Andreev band structure may or may not have a finite Chern number in the ground state.  

\begin{figure}
\includegraphics[width=\columnwidth]{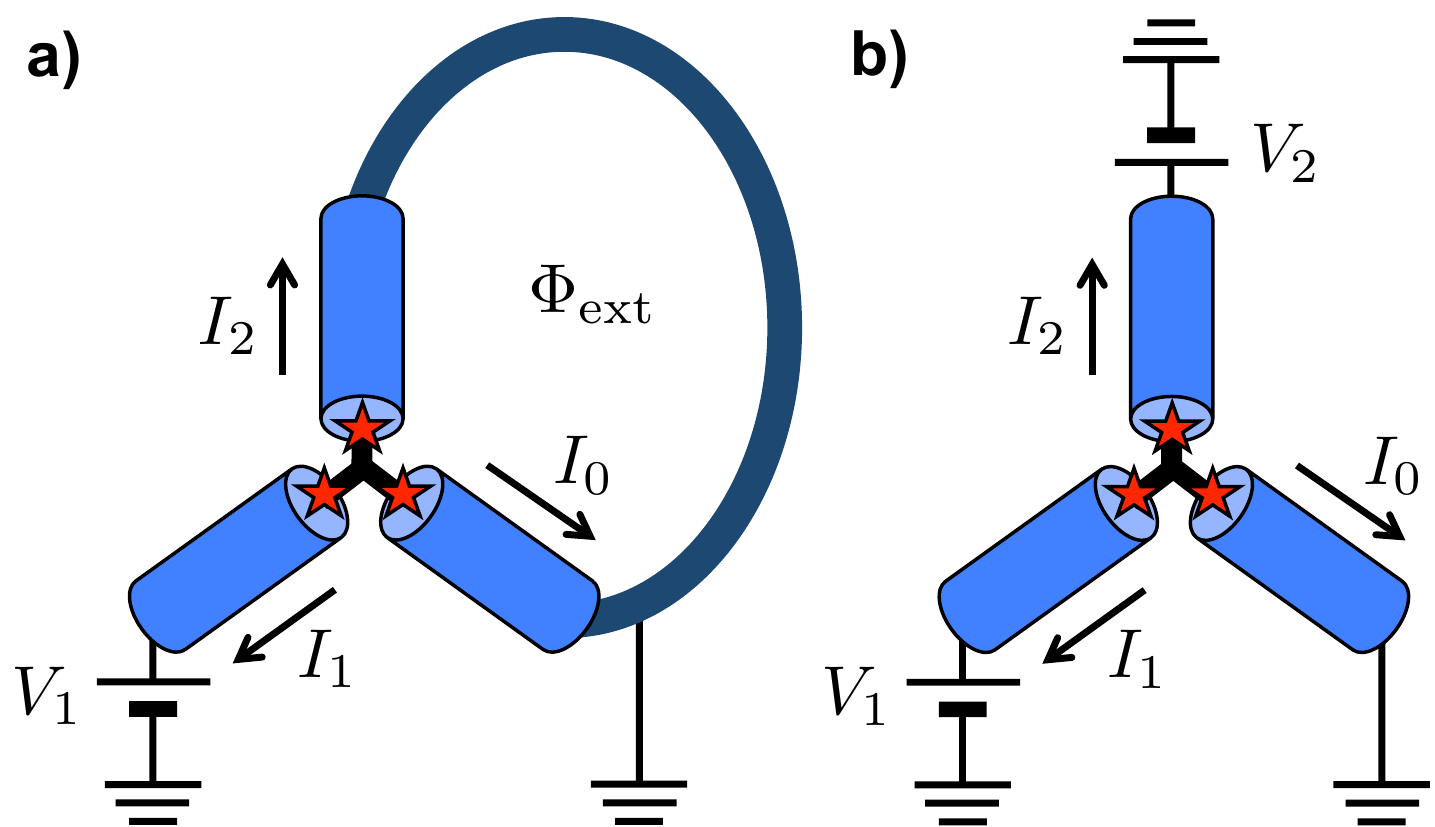}
\caption{
\label{fig:setup} 
Sketch of a trijunction made of three topological superconductors hosting Majorana modes at their ends. {{\bf a)} The superconducting phase difference $\phi_2$ between two of the terminals can be tuned by a magnetic flux $\Phi_\text{ext}$ applied through the superconducting loop that connects them. A voltage $V_1$ is applied to the remaining terminal. The transconductance  $G_{21}={\partial I_2}/{\partial V_1}$ is predicted to be quantized in units of $2e^2/h$ after averaging over $\phi_2$. {\bf b)} Alternatively, one can measure the transconductance between two voltage-biased terminals.}}
\end{figure} 

The analogy goes further. As in the case of real materials, a finite Chern number in the ground state is associated with a quantum Hall effect~\cite{TKKN}. In a multi-terminal junction, it manifests itself through the quantization  of the transconductance $G_{ij}={\partial I_i}/{\partial V_j}$, where $I_i$ is the {dc}  current in terminal $i$ and $V_j$ is the {dc} voltage applied to terminal $j\neq i$, when the entire space of the two superconducting phase differences is explored~\cite{Riwar2016}. Namely, one finds $G_{ij}=(2e^2/h) {g_s}C$, where $g_s$ is the spin degeneracy and $C$ is the sum of Chern numbers over all the occupied states. Due to the presence of a Fermi sea in the case of real materials, the exploration of the entire Brillouin zone is automatic. Here the Josephson relation ensures that the phase at the {v}oltage-biased terminal is swept linearly with time, whereas the second phase difference can be controlled using a magnetic flux through a loop connecting two of the terminals or, alternatively, with an additional voltage (see Fig.~\ref{fig:setup}). Thus, the transconductance probes the topological properties of the junction. Previous works showed that{, in certain parameter regimes,} a finite transconductance corresponding to a topologically non-trivial state may be obtained in junctions made with conventional superconductors (${g_s}=2$)~\cite{Riwar2016,Yokoyama2015,Eriksson2017,Meyer2017,Levchenko2017} and { with topological superconductors (${g_s}=1$)~\cite{Houzet2019,Gavensky2019}. In particular, r}eference~\cite{Gavensky2019} focussed on the fate of Weyl crossings across a topological transition in the leads {and showed the existence of a finite Chern number in the topological regime} in a specific trijunction setup.

Here we show that trijunctions made with topological superconductors {\em generically} have { a finite Chern number resulting in} a non-vanishing quantized transconductance, i.e., they are topologically non-trivial. The setup shown in Fig.~\ref{fig:setup} is the same as the one proposed for braiding Majorana modes in trijunctions and therefore of great interest in the context of topologically-protected quantum computation~\cite{Nayak2008}. 

A simple picture may be obtained by considering the low-energy sector only, which is described by the coupling of the three Majorana modes present at the junction. While two of the Majorana modes may couple to form a finite-energy Andreev state, one Majorana mode will always remain uncoupled and therefore yield a zero-energy state. We thus obtain a flat band at zero energy{~\cite{Alicea2011,Deb2018,Xie2019,Halperin2012,Weithofer2014,Zhou2014,Spanslatt2017},} plus a particle-hole conjugated state at energy $E_\pm=\pm |E_+|$. As we will show below, the latter crosses zero energy only if, { accidentally,} the system possesses {an effective (orbital) time-reversal symmetry (${\cal T}^2=1$, class BDI~\cite{BDI})}, such that { at some value of the superconducting phase differences} all three Majorana modes decouple. If { that symmetry is broken, as it is the case generically,} a gap exists at all values of the phases and {the ground state} carries a finite Chern number. The Chern number changes sign at the { effective} time-reversal invariant points associated with a three-band crossing (rather than a Weyl point). As a consequence, the junction displays a quantized transconductance $G_{ij}=\pm2e^2/h$. This signature is distinct from the case of trivial superconducting leads by two aspects: the transconductance is always finite, and the quantum of transconductance is {halved} due to the absence of spin degeneracy in the topological leads. 

Using numerical calculations of the current-voltage characteristics for representative junctions, we further show that an experimental detection of the transconductance quantization is well within reach.

\section{The model and main results}
\label{sec:II}

\subsection{Scattering theory and the effective Hamiltonian}

Our starting point to derive {these results} is a general formulation of the scattering problem describing the trijunction between {one-dimensional spinless $p$-wave superconductors. { (An effective low-energy model is presented in Appendix~\ref{appendix-tunnel}.)} Such topological leads} can be realized effectively with semiconducting nanowires and conventional superconductors using the proximity effect~\cite{Lutchyn2010, Oreg2010}. The equation determining the Andreev spectrum reads~\cite{Beenakker1991}
\begin{equation}
\label{eq-det}
{\rm det}\left[S(\epsilon)e^{i\hat \phi}S^*(-\epsilon)e^{-i\hat\phi}+a^{-2}(\epsilon)\mathbbm{1}\right]=0.
\end{equation}
Here $S(\epsilon)$ is the $3\times3$ scattering matrix of the junction in the normal state at energy $\epsilon$, $\hat\phi={\rm diag}[0,\phi_1,\phi_2]$ contains the superconducting phases of the terminals, $a(\epsilon)=\epsilon/\Delta-i\sqrt{1-(\epsilon/\Delta)^2}$ is the Andreev reflection amplitude, and $\Delta$ is the superconducting gap amplitude in the leads{~\cite{footnote}}.
The $p$-wave nature of the topological superconducting terminals is encoded in the $+$~sign between the two terms under the determinant~\cite{Houzet2019}, accounting for the phase shift of $\pi$ between the Andreev reflections for electrons and holes.

The eigenvalues of $\hat A=Se^{i\hat \phi}S^*e^{-i\hat\phi}$ are given as $\lambda_i=e^{i\alpha_i}$ with $\alpha_0=0$ and $\alpha_-=-\alpha_+$. Furthermore $E_i=\Delta\sin(\alpha_i/2)$. Thus, the eigenvalue $\lambda_0$ describes the flat band. For the Andreev states corresponding to $\lambda_\pm$ to cross the Fermi level, those eigenvalues have to be equal to 1 as well, i.e., $\hat A$ has to be the identity matrix. This requires $\Sigma^T=\Sigma$ with $\Sigma=e^{-i\hat\phi/2}S_0e^{i\hat\phi/2}$ and $S_0=S(\epsilon=0)$.
Inspired by the parametrization of the Cabibbo-Kobayashi-Maskawa quark mixing matrix, we find that the most general form of the scattering matrix can be written as
\begin{equation}
\!\!\!S_0=\begin{pmatrix}c_1&s_1c_3&s_1s_3\\s_1c_2&-c_1c_2c_3+s_2s_3 e^{i\delta}&-c_1c_2s_3-s_2c_3 e^{i\delta}\\
s_1s_2&-c_1s_2c_3-c_2s_3e^{i\delta}&-c_1s_2s_3+c_2c_3e^{i\delta}\end{pmatrix}
\end{equation}
with $c_i=\cos\theta_i$ and $s_i=\sin\theta_i$, and $\theta_i\in[0,\pi/2]$ and $\delta\in[0,2\pi[$, up to irrelevant phase factors that can be absorbed into $\phi_1,\phi_2$.
Then $\Sigma^T=\Sigma$ requires $\phi_1=\phi_2=0$ and $\theta_3=\theta_2$. This point is effectively time-reversal invariant. { At that point, the normal state conductances $G_{ij}^N=(e^2/h)(\delta_{ij}-|(S_0)_{ij}|^2)$ obey the relation $G_{ij}^N=G_{ji}^N$. For general parameters, we may define a \lq\lq chirality\rq\rq\, $\delta G=G^N_{ii+1}-G^N_{i+1i}=-(e^2/h)\sin^2\theta_1(\cos^2\theta_3-\cos^2\theta_2)$.}

To study the Chern number when { effective} time-reversal symmetry is broken, we may expand the determinant equation around the time-reversal invariant points, similar to Refs.~\cite{Yokoyama2015,Meyer2017}, to obtain an effective low-energy Hamiltonian at $|\phi_1],|\phi_2|,|\phi_3|\ll 1$ with $\phi_3\equiv \theta_3-\theta_2${, which describes} the three-band crossing. In particular, ${\cal H}\psi=\epsilon\psi$, where the effective Hamiltonian is 
\begin{equation}
\label{eq-H}
{\cal H}={i}\frac \Delta2\sum_{i=1}^3 \phi_i \left[\left(\bar\Sigma^*\right)^{1/2}
\overline{\frac{\partial\Sigma}{\partial\phi_i}}\left(\bar\Sigma^*\right)^{1/2}-\text{c.c.}\right].
\end{equation}
{ (For more details, see Appendix~\ref{appendix-Heff}.)} Here the bar denotes quantities evaluated at the { effective} time-reversal invariant point parametrized by $\theta_1,\theta_t=(\theta_2+\theta_3)/2,\delta$.
Using $\overline{\Sigma}=\overline{\Sigma}^T$ and $\overline{\partial_{\phi_i}\Sigma}=-\overline{\partial_{\phi_i}\Sigma}^T$, we find that $\cal H$ is antisymmetric. It is therefore equivalent to the Hamiltonian that describes a spin-1 in a magnetic field, ${\cal H}={\bf B}\cdot{\bf S}$, with magnetic field ${\bf B}=(B_x,B_y,B_z)$ and spin matrix vector ${\bf S}=(S_x,S_y,S_z)$. In particular, we obtain
\begin{equation}
\!\!\!\!\!\left(\!\begin{array}{c}B_x\\B_y\\B_z\end{array}\!\right)
\!=\!\frac\Delta 2\!
\left(\!\begin{array}{c}
-s_1(c_t^2\phi_1-s_{t}^2\phi_2)\\
s_1[s_{t}c_{t}d_+(\phi_1\!-\!\phi_2)-d_-\phi_3]\\
\!s_{t}c_{t}(d_-\!+\!c_1d_+)(\phi_1\!-\!\phi_2)\!+\!(d_+\!-\!c_1d_-)\phi_3\!
\end{array}\!\right)
\end{equation}
with $d_\pm=\cos(\delta/2)\pm\sin(\delta/2)$ and
\begin{eqnarray*}
S_x&=&i\left(\begin{array}{ccc}0&c_t&s_t\\-c_t&0&0\\-s_t&0&0\end{array}\right),
\quad S_y=i\left(\begin{array}{ccc}0&s_t&-c_t\\-s_t&0&0\\c_t&0&0\end{array}\right),\\
S_z&=&\left(\begin{array}{ccc}0&0&0\\0&0&i\\0&-i&0\end{array}\right).
\end{eqnarray*}
{We} see that the crossing of the eigenstates of  $\cal H$ is linear in the three parameters $\phi_1,\phi_2,\phi_3$. Furthermore, the Jacobian $J>0$ of the transformation from variables $B_{x,y,z}$ to $\phi_{1,2,3}$ can be used to show that a monopole with topological charge  $n_-={-2}$ ($n_+={+2}$) is associated with the state at negative (positive) energy~\cite{Berry1984}. 

\subsection{Chern numbers and the quantized transconductance}

We are now ready to define Chern numbers in the $(\phi_1,\phi_2)$-plane for each band. Using particle-hole symmetry, we conclude that the Chern number of the flat band is $C_0=0$, while the Chern number of the positive and negative energy bands are opposite and have a jump {from} $C_\pm={\mp 1}$ to $C_\pm={\pm 1}$ as $\phi_3$ is tuned across 0. Interestingly, the sign of the Chern numbers can be related to the sign of the \lq\lq chirality\rq\rq{,} { $\delta G=-{(e^2/h)}\sin^2\theta_1\sin(2\theta_t)\sin(\phi_3)$.} Namely,
\begin{equation}
C_\pm{=\mp{\rm sign}(\phi_3)}=\pm{\rm sign}(\delta G).
\end{equation}
Using adiabatic perturbation theory~\cite{Thouless1983,Riwar2016,Houzet2019}{, one can relate the transconductances $G_{12}$ and $G_{21}$ to the Chern number. (Details of the calculation are given in the Appendix of Ref.~\cite{Houzet2019}.) Assuming} that the system remains in the ground state (i.e., the state at negative energy is occupied whereas the state at positive energy is empty), we find the transconductances, 
\begin{equation}
G_{21}=-G_{12}=\frac{2e^2}h C_-.
\end{equation}
This is our main result announced above.

{ By relating the transconductance with the Chern number of the Andreev state near the zero-energy crossing, we rely on the established result that  the transconductance has to be zero when the system possesses { an effective (orbital)} time-reversal invariance and that it can change only due to zero-energy crossings between Andreev states: possible Weyl crossings at finite energy (between Andreev states or between an Andreev state and the continuum~\cite{footnote2}) do not affect the transport response, see e.g.~\cite{Riwar2016} as well as Fig.~6 in \cite{Gavensky2019} for a nice illustration. This ensures that our result relies on the low-energy properties of the system only and applies to more general models of topological trijunctions as well.}

\begin{figure}
\includegraphics[width=0.9\columnwidth]{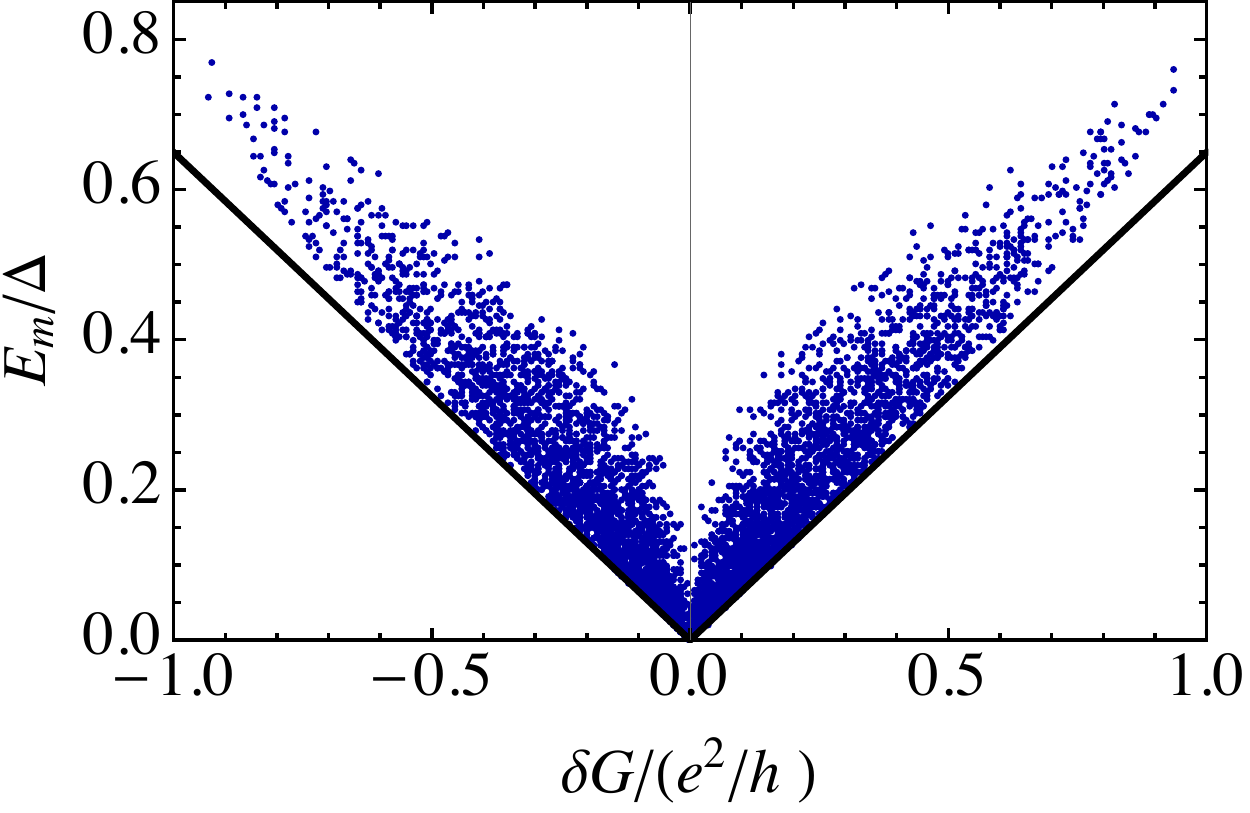}
\caption{
\label{fig:histogram} 
Value of the gap  $E_m=\min_{\phi_1,\phi_2}E_+$ as a function of the \lq\lq chirality\rq\rq\ $\delta G=G^N_{ii+1}-G^N_{i+1i}$ for {5,000} random scattering matrices drawn from CUE. {It illustrates the correlation between these two quantities, $E_m/\Delta \geq (3\sqrt3/8)|\delta G|/(e^2/h)$ {(indicated by the black line).}}} 
\end{figure} 

\section{Out-of-equilibrium effects and specific examples}
\label{sec:III}

Observing the conductance quantization requires applying a bias voltage{. T}herefore the adiabatic assumption used to obtain the above results has to be carefully analyzed{, as discussed in Ref.~\cite{Riwar2016}. On one hand, the voltage should be small on the scale of the superconducting gap, $eV\ll\Delta$. On the other hand, t}he voltage-induced dynamics of the superconducting phase may induce transitions between different states and, as a consequence, result in a non-equilibrium occupation of the Andreev spectrum. Parity-conserving transitions between the flat band and the finite-energy Andreev state tend to establish an equiprobability for each state to be occupied or empty, and therefore suppress the transconductance. Parity-changing transitions involving continuum states may be induced either by the presence of subgap states in the leads or by non-adiabatic processes. They tend to {empty (fill)  positive-energy (negative-energy) states},
and thus restore the transconductance quantization. The conditions for parity-changing transitions to dominate over the parity-conserving ones imposes constraints
on the voltage that can be applied, which are the more stringent the smaller the {gap}, $E_m=\min_{\phi_1,\phi_2}E_+$, in the Andreev spectrum.

\subsection{{ $I$-$V$}-characteristics for random scattering matrices}
\label{main-RMT}

Using random scattering matrices drawn from the circular unitary ensemble (CUE), which describes systems with broken { (orbital)} time-reversal symmetry, and taking the short junction limit, where the energy dependence of $S(\epsilon)$ can be neglected on the scale of $\Delta$, we find that the gaps are generically quite large (on the scale of $\Delta$){~\cite{footnote}}. Statistics are shown in Fig.~\ref{fig:histogram}. They indicate that the gap is correlated with the amount of time-reversal symmetry breaking encoded in the \lq\lq chirality\rq\rq\ $\delta G$, such that $E_m/\Delta \geq (3\sqrt3/8)|\delta G|/{(e^2/h)}${, where the numerical factor is derived using the effective Hamiltonian.}

\begin{figure}
\includegraphics[width=0.95\columnwidth]{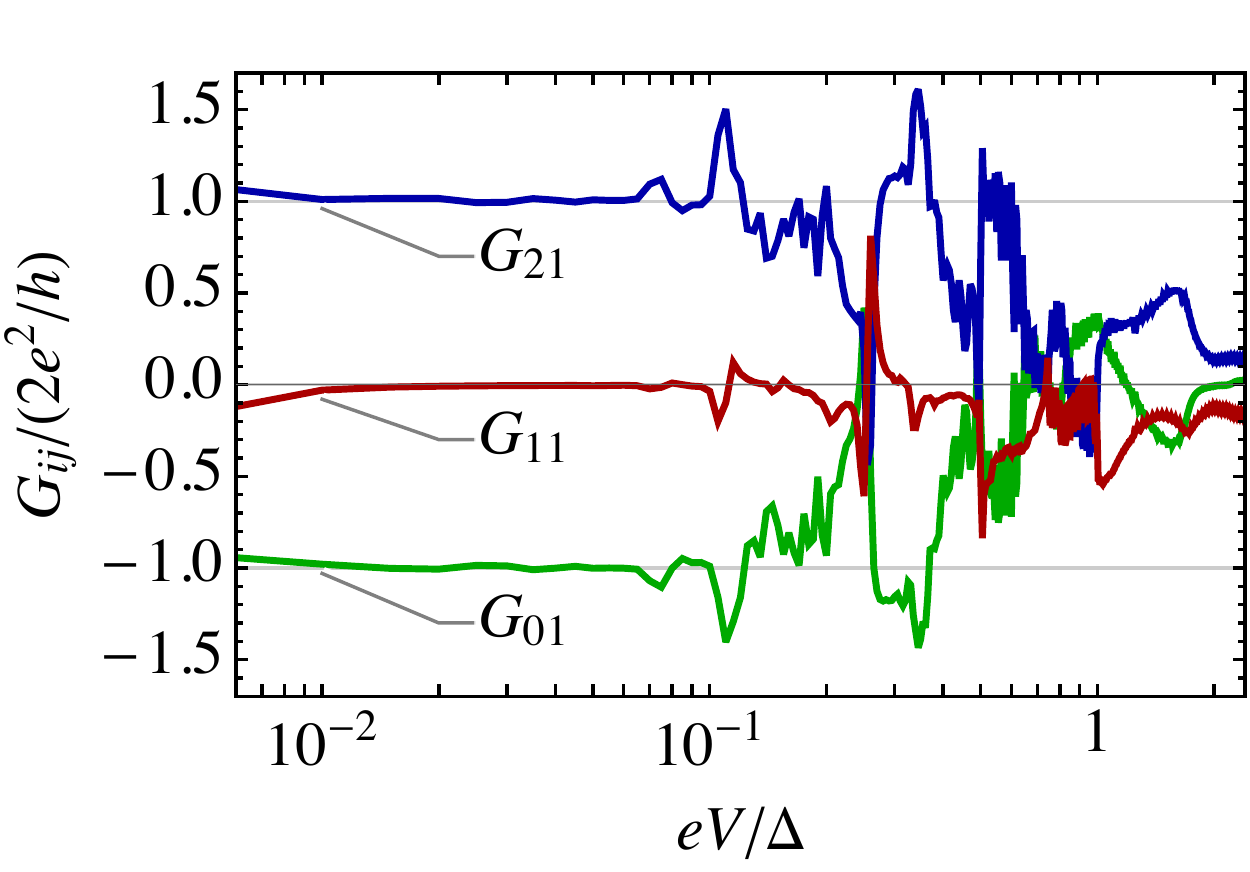}
\caption{
\label{fig:voltage} 
Differential conductance $G_{ij}=\partial I_i/\partial V_j$ as a function of voltage (log-scale) with a Dynes parameter $\gamma=2\times10^{-3}\Delta$ for a scattering {matrix} drawn from CUE with $\min_{\phi_1,\phi_2}E_+\approx0.37\Delta$, ${\delta G}\approx -0.26 e^2/h$. {The transconductance is quantized in an intermediate voltage regime. At very low voltages, the flat band induces dissipation, whereas, at high voltages, multiple Andreev reflections~\cite{Averin1995,Bratus1995} produce an intricate pattern of subgap features. Finally, at voltages $eV\gg\Delta$, the normal state conductances are recovered.} }
\end{figure}

{In order to check the robustness of the transconductance quantization beyond the adiabatic regime, we used} the formalism of multiple Andreev reflections~\cite{Eriksson2017,Averin1995,Bratus1995} {to compute} the current-voltage characteristics of various junctions with a gap in the range $(0.35-0.4)\Delta$. An example, corresponding
to the scattering matrix 
\begin{eqnarray*}
{ 
S_0\!=\!\!\begin{pmatrix}0.628 \!+\! i 0.118&
  0.187 \!- \!i 0.056&
  0.743 \!- \!i
   0.044\\
   0.521\! - \!
   i 0.151 & -0.724\! +\! i
   0.325& -0.198 \!+\! i
   0.195\\-0.143\!+ \!
   i 0.526&-0.147\! + \!
   i 0.558&
  0.082 \!-\! i 0.602\end{pmatrix}\!\!,}
  \end{eqnarray*}
 is shown in Fig.~\ref{fig:voltage} (for more examples, see Appendix~\ref{appendix-rmt}). 
 
 To account for inelastic relaxation processes, we used a Dynes parameter~\cite{dynes} to smear the density of states in the terminals. {While the flat band does not contribute to the quantized value of the adiabatic transconductance, it may lead to large dissipation at small voltages in the presence of a finite Dynes parameter in the superconducting leads}. Namely, if the effective density of states in the leads is finite below the gap edge, one probes the zero-bias peak associated with the Majorana mode. For example, it can be seen in Fig.~\ref{fig:voltage} that $G_{11}$ vanishes over a range of voltages and then starts deviating from zero again as voltage is further decreased. Thus, the expected conductance quantization is seen in an intermediate voltage range.
 { This is similar to the condition of an  { intermediate time scale} required for Majorana braiding in trijunctions~\cite{Meng2011}. Indeed, in that case the manipulation should be { slow on the time scale set by the bulk gap to avoid exciting the system, but fast on the time scale set by the splitting between the different states for them to appear degenerate.}}

A small but finite temperature, $T\ll\Delta$, would smear the zero-bias peak and therefore may help to restore the conductance quantization. {For more details on the effect of the Dynes parameter and temperature, see Appendix~\ref{appendix-Dynes}.} {Though} the flat band does not affect the transconductance, except at very low voltage, switchings in its occupation result in a giant shot noise, as recently predicted in the case of an { effectively} time-reversal invariant junction~\cite{Jonckheere2019}. This might affect the averaging time necessary to extract the quantized value{, as discussed in the {Supplementary Note 3} of Ref.}~\cite{Riwar2016}.

\subsection{Symmetric junction and the role of time-reversal symmetry breaking}
\label{main-symmetric}

\begin{figure}
\includegraphics[width=0.95\columnwidth]{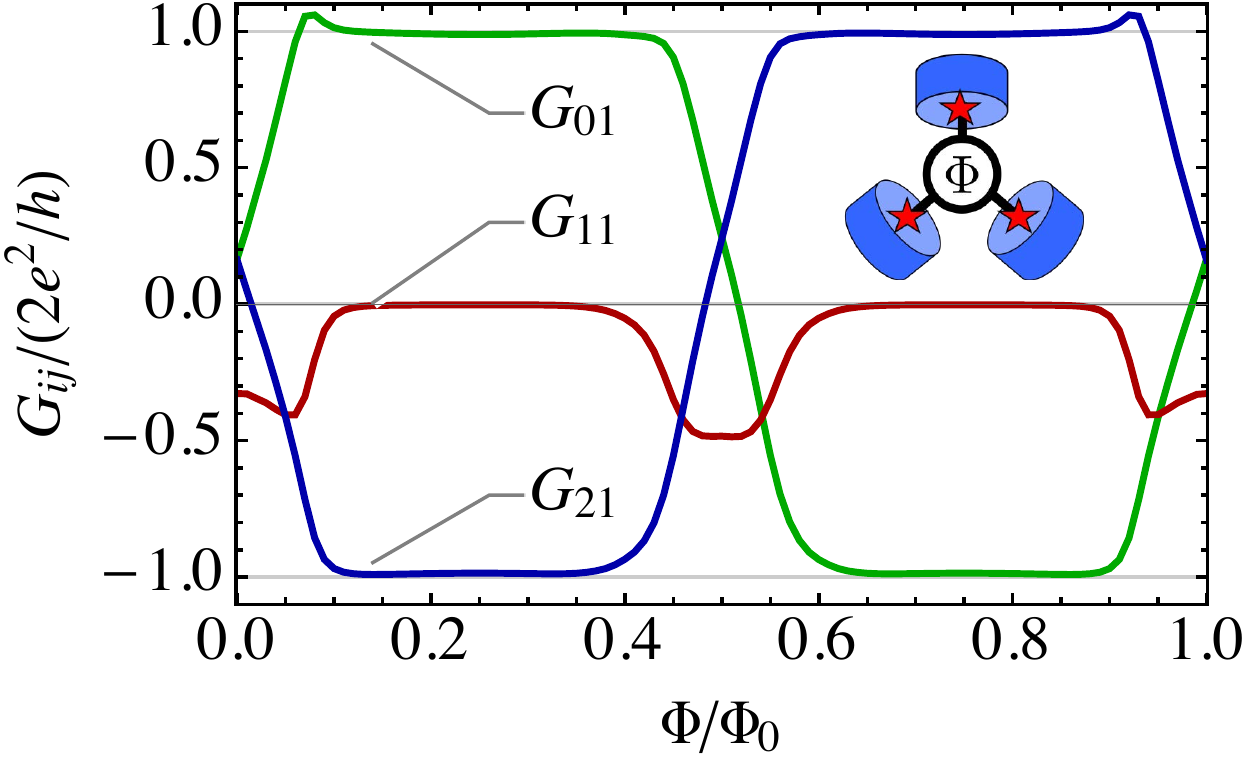}
\caption{
\label{fig:quantized} 
Conductance as a function of flux {$\Phi$ through the junction area} of a symmetric trijunction {(see the inset for a sketch of its central region)} {with $U=0$, $t=1/\pi\nu$, and} a Dynes parameter $\gamma=10^{-3}\Delta$ at voltage $eV=0.1\Delta$. {Plateaus of quantized transconductance are visible away from time-reversal invariant values of the flux, $\Phi/\Phi_0=0,1/2$, while a large dissipation is generated in their vicinity. 
}}
\end{figure}

To better show the role of time-reversal symmetry breaking, we also studied a toy model of a symmetric junction subject to a magnetic flux $\Phi$ {through the junction area (not to be confused with the flux $\Phi_{\rm ext}$ shown in Fig.~\ref{fig:setup} used to tune the superconducting phase difference)}. In particular,
we model a junction that consists of superconducting leads that are connected via a single site with on-site energy {$U$ and a hopping matrix element $t$ between the leads with normal density of states $\nu$}~\cite{Meyer2017}. The corresponding scattering matrix reads $S_0=(1-i\pi\nu W)^{-1}(1+i\pi\nu W)$, where
 \begin{equation}
 W=\begin{pmatrix}U&te^{is/3}&te^{-is/3}\\te^{-is/3}&U&te^{is/3}\\
 {te^{is/3}}&te^{-is/3}&U\end{pmatrix}
 \end{equation}
is the coupling matrix and the parameter $s=2\pi\Phi/\Phi_0$ describes the magnetic flux through the junction.
The conductance as a function of flux at low voltage is shown in Fig.~\ref{fig:quantized}. The change of sign of the transconductance at the { effectively} time-reversal invariant points, $\Phi/\Phi_0=0,1/2$, {where $\Phi_0$ is the flux quantum}, can be clearly seen along with an increase of $G_{11}$ showing the increased dissipation due to the closing of the gap. 

Within the same model, we can also study the effect of the Majorana modes at the far 
 end of the wire: as long as the coupling is sufficiently weak such that the splitting of the flat band is much smaller than the gap to the finite-energy Andreev {s}tate, the conductance quantization survives (see Appendix~\ref{appendix-4M}). As shown in Ref.~\cite{Houzet2019}, if the coupling becomes sufficiently strong, { Majorana-}Weyl crossings at finite energy may render the junction trivial.
{ {Furthermore,  a} generalization of the model introduced in Appendix~\ref{appendix-ABS} allows  {one} to illustrate the loss of the transconductance quantization when the Majorana end modes of two topological superconducting leads are hybridized with an accidental zero-energy Andreev bound state at the end of a conventional superconducting {lead}. Thus our prediction clearly distinguishes an isolated Majorana mode from an accidental zero-energy Andreev bound state.}

\section{Conclusion}

In conclusion, we have shown that Josephson trijunctions made with topological superconductors, as the ones proposed to braid Majorana modes, generically display a robust quantization of the transconductance that allows one to distinguish them from the case where the superconductors are topologically trivial. Our findings thus provide a novel signature of the presence of Majorana modes in such junctions. { The experimental observation of the transconductance quantization proposed here could be used as an important milestone on the way to Majorana braiding.}

\begin{acknowledgments}
We thank G.~Usaj for sending us the preprint of Ref.~\cite{Gavensky2019} and A.~Levy Yeyati for an interesting discussion about the noise in such junctions. Furthermore, we acknowledge funding by the ANR through the grant ANR-17-PIRE-0001.
\end{acknowledgments}

\begin{widetext}
\begin{appendix}

{

{\section{Effective low-energy theory for weakly coupled leads}
\label{appendix-tunnel}

A simple model can be derived in the case of a tunnel junction made of three weakly coupled one-dimensional spinless $p$-wave superconductors, similar to what was done in the 4-terminal case~\cite{Houzet2019}. In that case, the effective low-energy Hamiltonian  in terms of the three Majorana end modes at the junction takes the form
\begin{equation}
\label{eq:Htunnel}
H=\frac i2\sum_{0\leq i<j\leq 2} \xi_{ij} \gamma_i\gamma_j
\end{equation}
with $\xi_{ij}= |t_{ij}|\sin\left[(\chi_i-\chi_j)/2-\phi_{ij}\right]$.
Here, $\gamma_i$ is a Majorana fermionic operator (such that $\gamma_i^2={1}$), which describes the Majorana zero mode at the end of superconductor $i$, with superconducting phase  $\chi_i$, when it is decoupled from the others. Furthermore, $t_{ij}=|t_{ij}| e^{i\phi_{ij}}=t_{ji}^*$ are proportional to the tunneling matrix elements for electrons between leads $i$ and $j$. 

Next, we introduce fermion operators, $c={(\gamma_1+i\gamma_2)/2}$ and $c_a={(\gamma_0+i\gamma_a)/2}$, where $\gamma_a$ is an auxiliary Majorana mode that does not appear in the Hamiltonian. Then, the Hamiltonian~\eqref{eq:Htunnel} reads $H=\frac 12 {\cal C}^\dagger {\cal H}{\cal C}$, where ${\cal C}=(c,-\gamma_0/\sqrt 2,-c^\dagger)^T$ and $
{\cal H}=\bm{B}\cdot\bm{S}$.
Here $\bm{S}=(S_x,S_y,S_z)$ is a vector of spin-1 matrices, 
\begin{equation}
S_x=\frac1{\sqrt2}\begin{pmatrix}0&1&0\\1&0&1\\0&1&0\end{pmatrix},\qquad S_y=\frac1{\sqrt2}\begin{pmatrix}0&-i&0\\i&0&-i\\0&i&0\end{pmatrix}, \qquad S_z=\begin{pmatrix}1&0&0\\0&0&0\\0&0&-1\end{pmatrix},
\end{equation}
and
$\bm{B}=(\xi_{20},\xi_{01},\xi_{12})^T$.
The eigenenergies are $E_0=0$ and $E_\pm=\pm|\bm{B}|$. The unitary transformation diagonalizing the Hamiltonian is given as $U=\exp[i\theta_zS_z]\exp[i\theta_yS_y]$ with $\tan\theta_z=\sqrt{\xi_{20}^2+\xi_{01}^2}/\xi_{01}$ and $\tan\theta_y=\xi_{01}/\xi_{20}$. The Majorana mode associated with the flat band is thus $\tilde\gamma_0=\cos\theta\gamma_0+\sin\theta[\cos\phi(c+c^\dagger)+i\sin\phi(c-c^\dagger)]$.

Thus $E_\pm$ is finite unless $\xi_{01}=\xi_{12}=\xi_{20}=0$. This condition can be fulfilled by adjusting the superconducting phases only when $\phi_{01}+\phi_{12}+\phi_{20}=n\pi$ with $n\in\mathbb{Z}$. This particular choice for the phases of the tunneling matrix elements corresponds to the effectively time-reversal invariant point discussed in the main text. In all other cases, the Andreev spectrum is gapped at any value of the superconducting phases. 

While the supercurrent carried by the Andreev states depends on the strength of the tunnel couplings, $|t_{ij}|$, the quantized transconductance does not as is a topological feature solely related to the behavior of the wavefunctions within the \lq\lq Brillouin zone\rq\rq defined by the superconducting phase difference.

If we use the Majorana mode at the far end of the terminal 0 as the auxiliary Majorana mode, $\langle c_a^\dagger c_a\rangle$ measures the occupation of the zero-energy state in terminal 0. It can be related to the occupation $n_{\rm flat}=(1+i\langle\tilde\gamma_0\gamma_a\rangle)/2$ of the flat band appearing in the Andreev spectrum as
\begin{equation}
n_0=\langle c_a^\dagger c_a\rangle=\frac12\left[1+\frac{\xi_{12}}{E_+}\left(2n_{\rm flat}-1\right)\right].
\end{equation}
We now turn to the evolution of that occupation as the superconducting phases are varied. Terminal 0 is decoupled at superconducting phases $\chi_1=-2\phi_{01},\chi_2=2\phi_{20}$. Without loss of generality, we may choose $\phi_{01}=\phi_{20}=0$ such that
\begin{equation}
n_0(0,0)=\frac12\left[1+{\rm sign}(\xi_{12}(0,0))\left(2n_{\rm flat}-1\right)\right].
\end{equation}
As we consider an adiabatic evolution, we may assume $n_{\rm flat}$ to be constant. For illustration, let us set $n_{\rm flat}=(1-{\rm sign}(\xi_{12}(0,0))/2$, which yields $n_0(0,0)=0$. If one of the phases is advanced by $2\pi$, one reaches a point where terminal 0 is again decoupled. As $\xi_{12}= |t_{12}|\sin\left[(\chi_1-\chi_2)/2-\phi_{12}\right]$ has switched sign in that process, the occupation has changed to $n_0(2\pi,0)=n_0(0,2\pi)=1$. If both phases are advanced by $2\pi$, one gets back to $n_0(2\pi,2\pi)=n_0(0,0)=0$. 

This universal result suggests a connection between the quantized transconductance and braiding. Braiding protocols based on flux-tuned tri-junctions have indeed been proposed~\cite{Frolov}. However, while braiding relies on the properties of the flat band, the transconductance probes the Chern number of the finite-energy Andreev band. A detailed analysis is beyond the scope of this work.
}

\section{Derivation of the effective Hamiltonian}
\label{appendix-Heff}

Close to the time-reversal symmetric point the determinant equation~\eqref{eq-det},
${\rm det}\left[\Sigma\Sigma^*+a^{-2}(\epsilon)\mathbbm{1}\right]=0$,
 may be written in the following form, expanding up to linear order in $\phi_i$ and $\epsilon$:
\begin{eqnarray}
{\rm det}\left[\sum_{i=1,2,3}\left(\overline{\frac{\partial\Sigma}{\partial\phi_i}}\bar\Sigma^*+\bar\Sigma\overline{\frac{\partial\Sigma^*}{\partial\phi_i}}\right)\phi_i-2i\frac\epsilon\Delta\mathbbm{1}\right]=0.
\end{eqnarray}
{Here we neglect the energy-dependence of the normal state scattering matrix $S(\epsilon)$ on the scale of the Thouless energy as it does not qualitatively change our conclusions.}
This allows one to define the effective Hamiltonian given in Eq.~\eqref{eq-H}. Taking the derivatives, the explicit form of the Hamiltonian reads
\begin{eqnarray}
{\cal H}
&=&\frac i2\Delta\Bigg\{\begin{pmatrix}0&-s_1(c_t^2-s_t^2d_+)&-s_1s_tc_t(1+d_+)\\
s_1(c_t^2-s_t^2d_+)&0&s_{t}(d_-+c_1d_+)\\
s_1s_tc_t(1+d_+)&-s_{t}(d_-+c_1d_+)&0
\end{pmatrix}c_t\phi_1\nonumber\\
&&\qquad+\begin{pmatrix}0&s_1s_tc_t(1-d_+)&s_1(s_t^2+c_t^2d_+)\\
-s_1s_tc_t(1-d_+)&0&-c_{t}(d_-\!+\!c_1d_+)\\
-s_1(s_t^2+c_t^2d_+)&c_{t}(d_-\!+\!c_1d_+)&0
\end{pmatrix}s_t\phi_2\nonumber\\
&&\qquad+\begin{pmatrix}0&-s_1s_td_-&s_1c_td_-\\
s_1s_td_-&0&d_+-c_1d_-\\
s_1c_td_-&-(d_+-c_1d_-)&0
\end{pmatrix}\phi_3\Bigg\}.
\end{eqnarray}
A more convenient form is given in the main text, where we write the antisymmetric Hamiltonian as ${\cal H}={\bf B}\cdot{\bf S}$, with magnetic field ${\bf B}=(B_x,B_y,B_z)$ and spin matrix vector ${\bf S}=(S_x,S_y,S_z)$. 
}

\section{Quantization of the transconductance for different scattering matrices}
\label{appendix-rmt}

In section~\ref{main-RMT} (Fig.~\ref{fig:voltage}), we presented numerical data for one random scattering matrix. In Fig.~\ref{fig:voltage-SM}, we show the results for several {representative} matrices with a gap in the range $(0.35-0.4)\Delta$ to illustrate the robustness of the transconductance quantization.

\begin{figure}[h]
(a) \includegraphics[width=0.275\columnwidth]{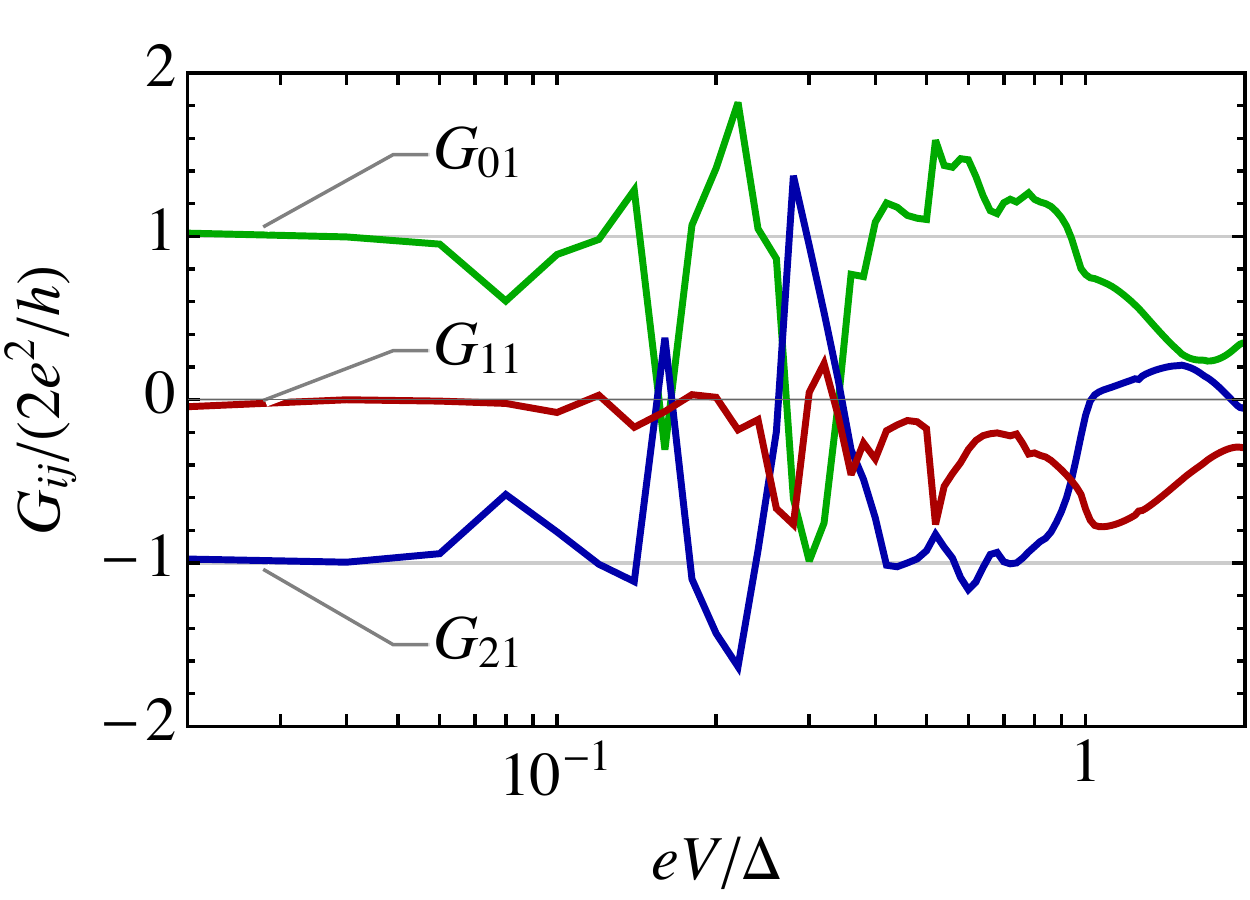}
(b) \includegraphics[width=0.275\columnwidth]{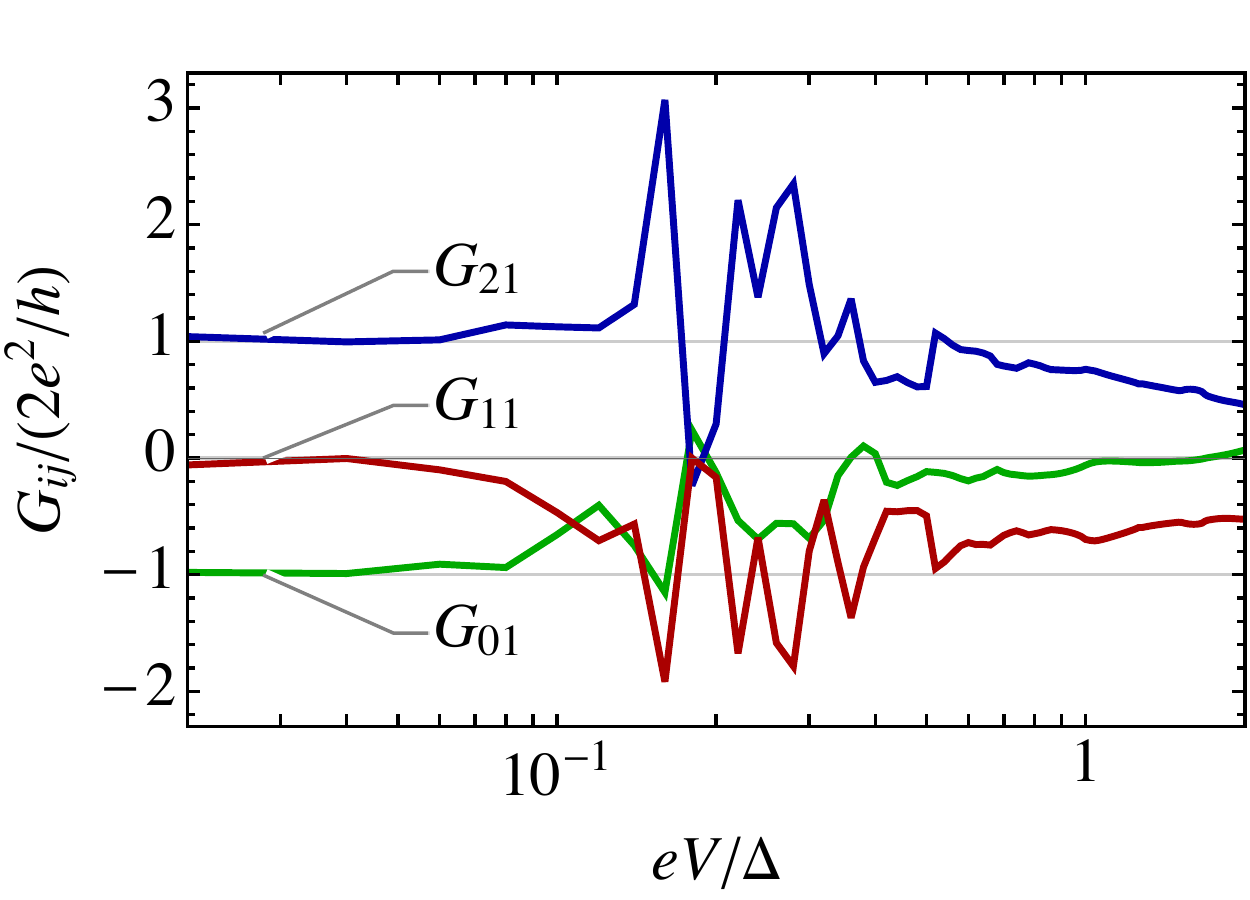}
(c) \includegraphics[width=0.275\columnwidth]{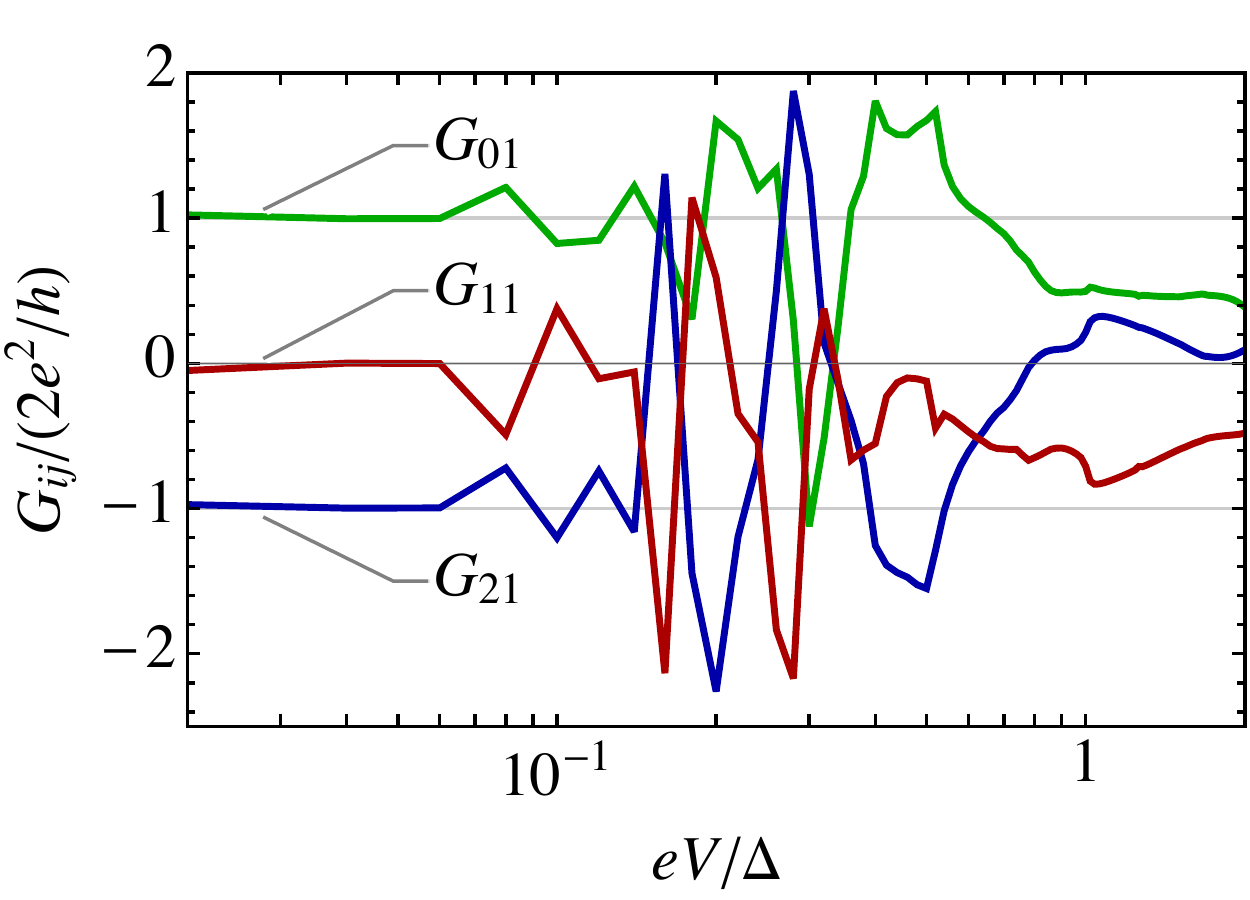}
\caption{
\label{fig:voltage-SM} 
Differential conductance $G_{ij}=\partial I_i/\partial V_j$ as a function of voltage for various random scattering matrices with a Dynes parameter $\gamma=10^{-3}\Delta$. (a) $\min_{\phi_i}E_+\approx0.35\Delta$, $ \delta G\approx 0.46 e^2/h$. (b) $\min_{\phi_i}E_+\approx0.40\Delta$, $\delta G\approx -0.34 e^2/h$. (c) $\min_{\phi_i}E_+\approx0.39\Delta$, $\delta G\approx 0.50 e^2/h$.}
\end{figure}

\section{Effect of the Dynes parameter and temperature}
\label{appendix-Dynes}

As shown in section~\ref{main-RMT} (Fig.~\ref{fig:voltage}), the flat band leads to deviations from the predicted conductance quantization at low voltages. The value of the voltage  {$V^*$}, where these deviations become visible, depends on the Dynes parameter. {Namely, $eV^*\sim\gamma$, corresponding to the effective width of the uncoupled Majorana mode yielding the flat band.} Using the same scattering matrix as in the main text, we show the voltage dependence of the conductances in the low-voltage regime for various values of the Dynes parameter in Fig.~\ref{fig:Dynes-SM}.

\begin{figure}[h]
\includegraphics[width=0.35\columnwidth]{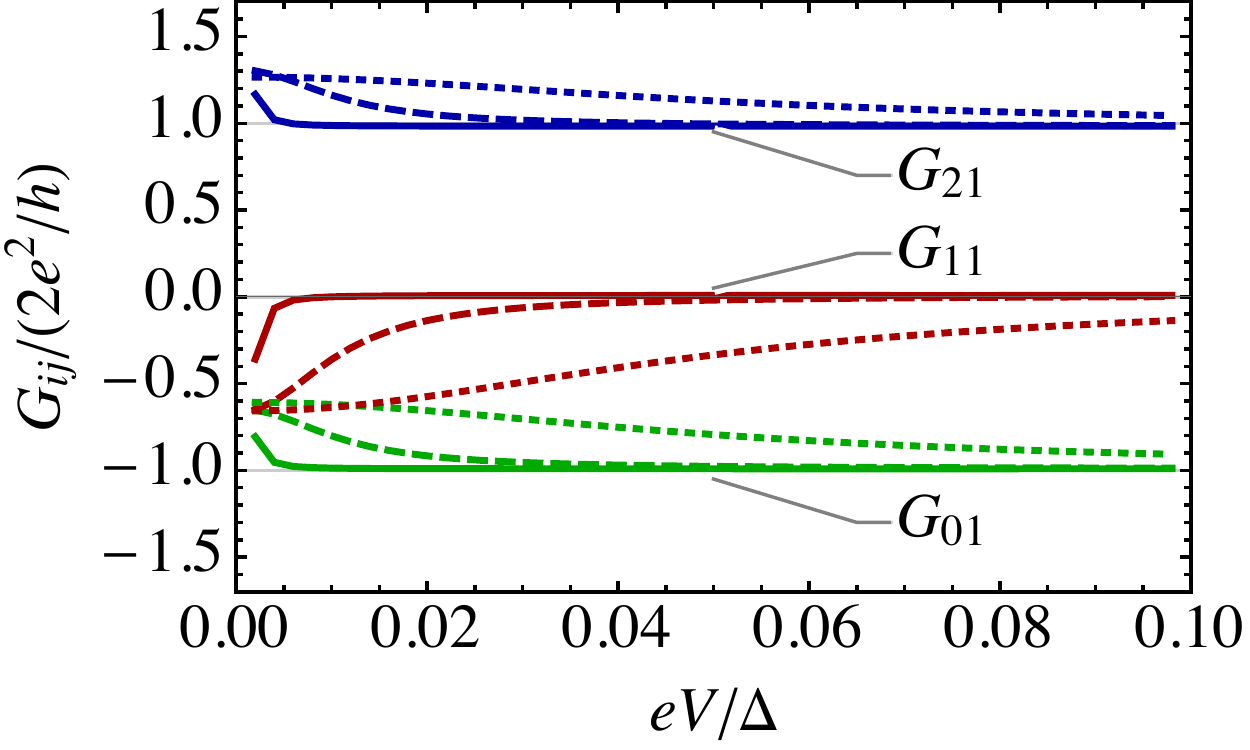}
\caption{
\label{fig:Dynes-SM} 
Differential conductance $G_{ij}=\partial I_i/\partial V_j$ for various values of the Dynes parameter $\gamma=0.001$ (full lines), $\gamma=0.01$ (dashed lines), and $\gamma=0.05$ (dotted lines). The larger $\gamma$, the larger the voltage at which deviations from conductance quantization appear in the low-voltage regime.}
\end{figure} 

It turns out that the conductance quantization at these low voltages improves with temperature. The results presented in the main text were obtained at zero temperature. It is straightforward to extend te calculations to finite temperature. As shown here, temperature suppresses the large dissipation due to the flat band at small voltages. In Fig.~\ref{fig:T_dep-SM} we show the dependence of the conductance on temperature at a fixed voltage. Conductance quantization is recovered when $T\gtrsim \gamma$.

\begin{figure}[h]
(a) \includegraphics[width=0.275\columnwidth]{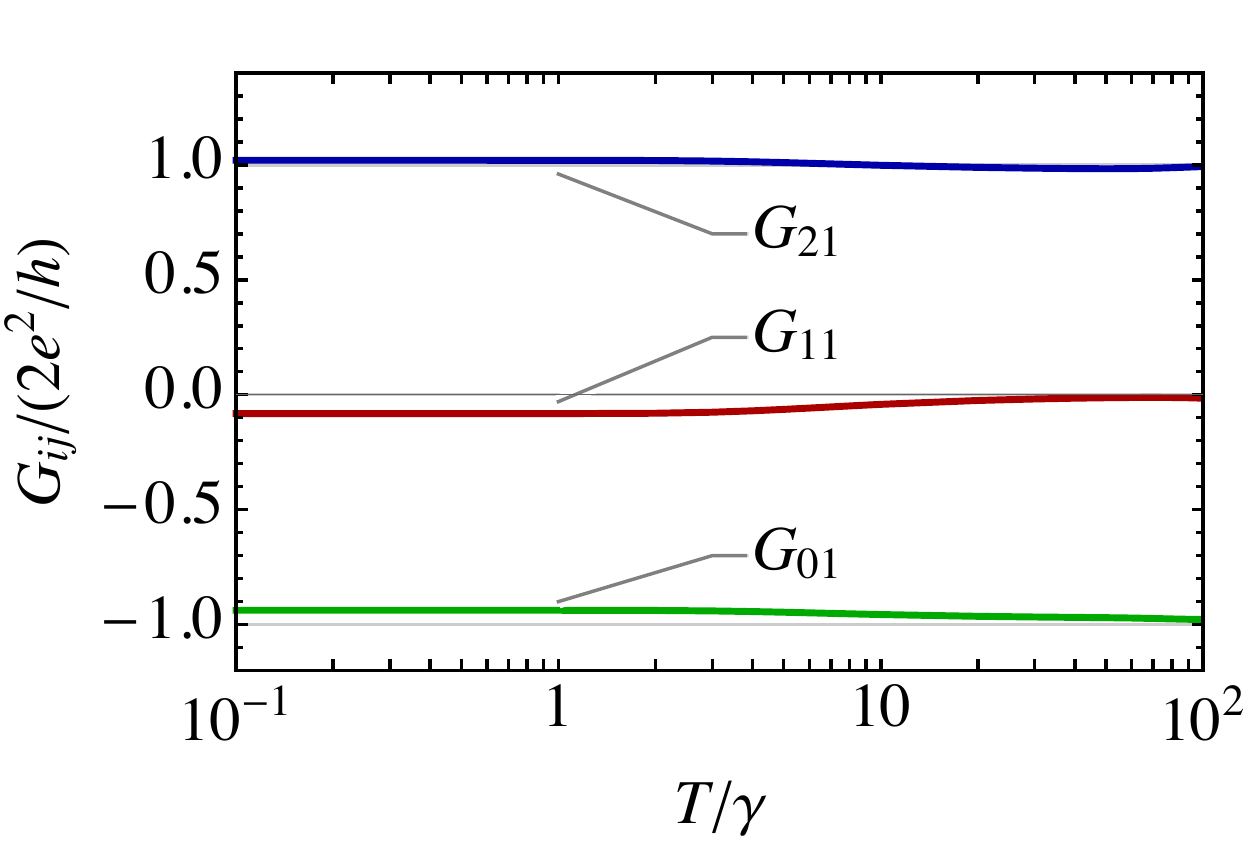}
(b) \includegraphics[width=0.275\columnwidth]{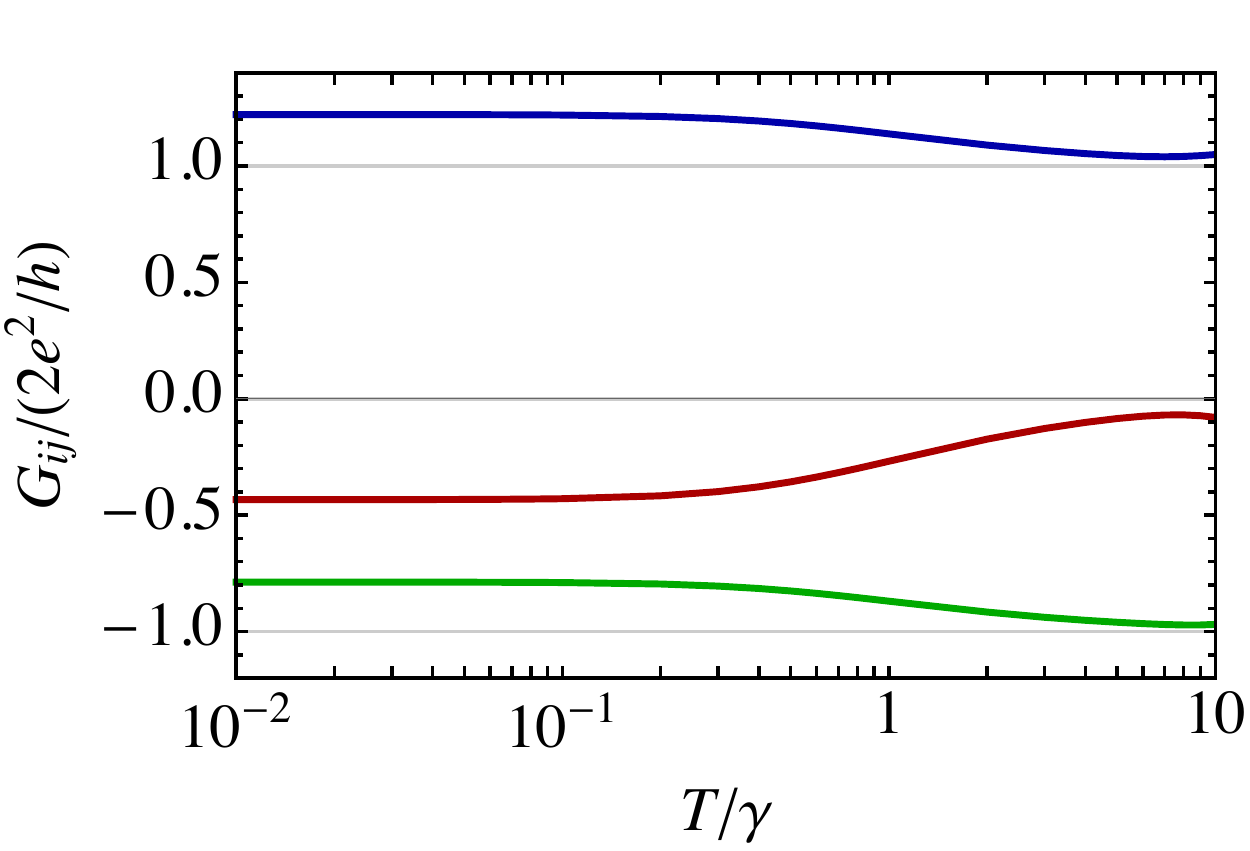}
(c) \includegraphics[width=0.275\columnwidth]{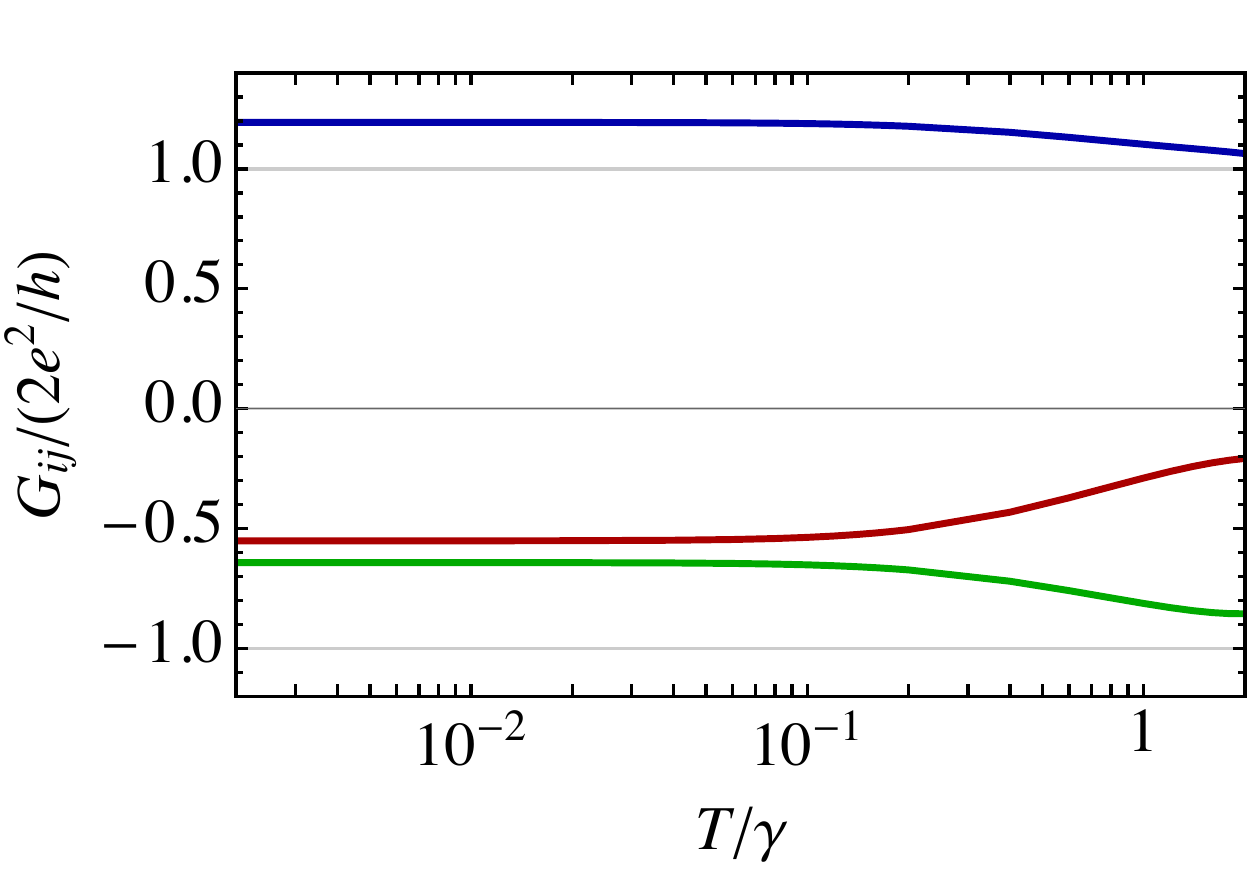}
\caption{
\label{fig:T_dep-SM} 
Temperature dependence of the conductance $G_{ij}=I_i/V_j$ at voltage $V=0.1\Delta/e$, and with a Dynes parameter (a) $\gamma=10^{-3}\Delta$, (b) $\gamma=10^{-2}\Delta$ and (c) $\gamma=5\times10^{-2}\Delta$.}
\end{figure} 

\section{Effect of a finite length of the topological superconductors}
\label{appendix-4M}

In our calculations, we only accounted for the Majorana modes localized at the junction, neglecting the Majorana modes localized at the far end of the wires. If the wires have a finite length, the two Majorana modes at either end may couple. Furthermore, inhomogeneities may lead to the appearance of additional Majorana modes at intermediate distances. Here we investigate how a coupling to an additional Majorana mode in one of the terminals modifies our prediction. 

As a starting point, we use a symmetric junction as described in section~\ref{main-symmetric} and add an additional channel to terminal 2. This additional channel is only coupled to the other channel in the same terminal via a hopping matrix element $t_0$, modeling the coupling between the two Majoranas at opposite ends of the wire, while no direct coupling to the other terminal exists. The coupling matrix $W$ then reads
\begin{equation}
W=\begin{pmatrix}U&te^{is/3}&te^{-is/3}&0\\te^{-is/3}&U&te^{is/3}&0\\
 {te^{is/3}}&te^{-is/3}&U&i t_0\\0&0&-it_0&0\end{pmatrix}.
 \end{equation}
For small $t_0$, the main effect of the coupling  to the additional Majorana is to modify the flat band {such} that it acquires a finite dispersion. However, as long as the gap to the finite-energy state remains sufficiently large, the conductance quantization remains robust. At larger $t_0$, the two states may cross, signaling a transition out of the topological regime such, that the conductance quantization breaks down. This is shown in Fig.~\ref{fig:3T4T-SM}.

\begin{figure}[h]
\includegraphics[width=0.35\columnwidth]{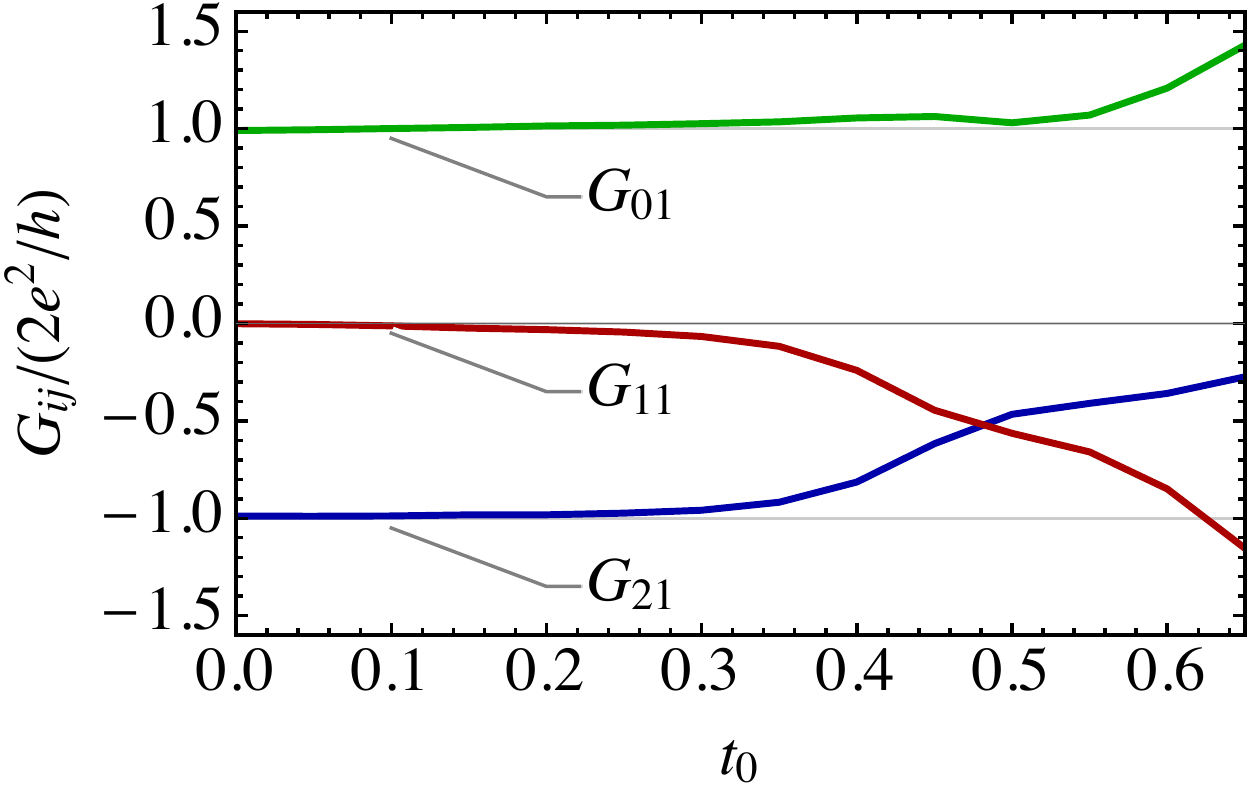}
\caption{
\label{fig:3T4T-SM} 
Conductance $G_{ij}=I_i/V_j$ as a function of the coupling $t_0$ at voltage $V=0.15\Delta$ and with a Dynes parameter $\gamma=10^{-3}\Delta$. Here $t=0.6$ and $U=0$.}
\end{figure}

{
\section{Accidental zero-energy Andreev bound states versus isolated Majorana modes}
\label{appendix-ABS}

A major question in studying topological junctions is how to distinguish isolated Majorana modes from accidental zero-energy Andreev bound states. To show that our prediction clearly distinguishes between the two, we consider here a situation where one of the superconducting leads is trivial, but hosts an accidental zero-energy Andreev bound state.

As in {A}ppendix \ref{appendix-4M}, we use a symmetric junction as described in section~\ref{main-symmetric} as a starting point. We add an additional channel to terminal 2 such that the two Majorana modes in terminal 2 form a zero-energy Andreev bound state. To achieve this, we assume that the two Majorana modes do not couple to each other whereas they are coupled to the Majorana modes in the other terminal via the hopping Hamiltonian
\begin{equation}
H_{2j}=\frac12t_{2j}c_2^\dagger\gamma_j+h.c.,
\end{equation}
where the fermionic creation operator $c_s^\dagger$ is formed by the two Majorana modes in terminal 2, namely, $c_2^\dagger=(\gamma_{2a}-i\gamma_{2b})/2$. In particular,
\begin{equation}
H_{2j}=\frac i2|t_{2j}|\sin\left(\frac{\chi_2-\chi_j}2-\phi_{2j}\right)\gamma_{2a}\gamma_j+\frac i2|t_{2j}|\sin\left(\frac{\chi_2-\chi_j}2-\phi_{2j}-\frac\pi2\right)\gamma_{2b}\gamma_j.
\end{equation}
To interpolate between the case of an isolated Majorana mode and an accidental zero-energy Andreev bound state, we multiply the second term by $\alpha$ and vary $\alpha$ from zero (isolated Majorana mode) to one (accidental zero-energy Andreev bound state).
For a symmetric junction, the coupling matrix $W$ then reads
\begin{equation}
W=\begin{pmatrix}U&te^{is/3}&te^{-is/3}&i\alpha te^{-is/3}\\te^{-is/3}&U&te^{is/3}&i\alpha te^{is/3}\\
 {te^{is/3}}&te^{-is/3}&U&0\\-i\alpha te^{-is/3}&-it\alpha e^{is/3}&0&U\end{pmatrix}.
\end{equation}
The conductances as a function of $\alpha$ are illustrated in Fig.~\ref{fig:ABS-SM}. One sees clearly that the quantization of the transconductance breaks down as $\alpha$ increases. When $\alpha\to{1}$, the conductance $G_{21}$ vanishes: as shown previously for two-terminal junctions{~\cite{Peng2015}}, no subgap current can flow between a conventional and a topological superconductor. The large direct conductance $G_{11}$ in that limit is due to the dispersing Andreev bound states involving the Majorana modes of terminals 0 and 1.
\begin{figure}[h]
\includegraphics[width=0.35\columnwidth]{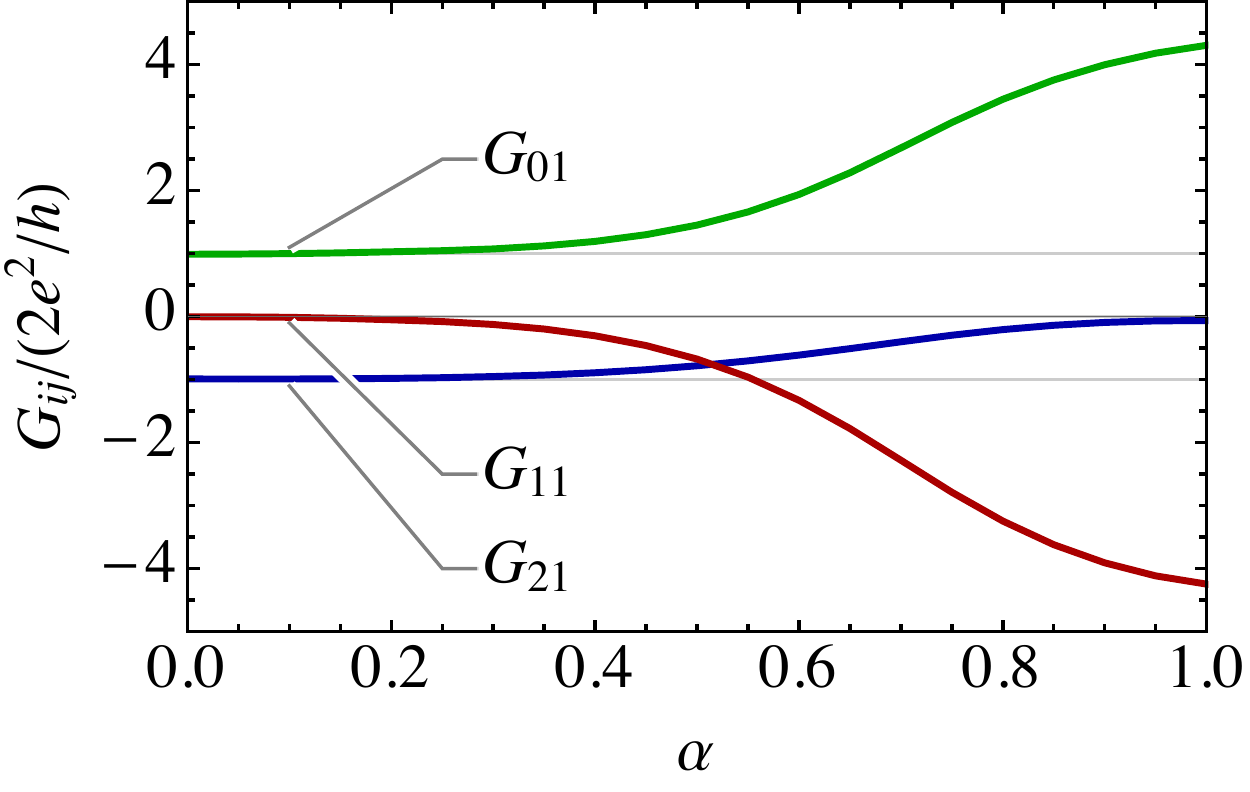}
\caption{
\label{fig:ABS-SM} 
Conductance $G_{ij}=I_i/V_j$ as a function of the coupling $\alpha$ at voltage $V=0.15\Delta$ and with a Dynes parameter $\gamma=10^{-3}\Delta$. Here $t=0.6$ and $U=0$.}
\end{figure} 

}

\end{appendix}
\end{widetext}

\end{document}